\documentclass[seceq]{ptptex}

\usepackage{graphicx}

\newcommand{\be}{\begin{equation}}
\newcommand{\ee}{\end{equation}}
\newcommand{\ba}{\begin{eqnarray}}
\newcommand{\ea}{\end{eqnarray}}

\newcommand{\De}{\Delta}

\title{Dynamically generated resonances
}

\author{ \scshape 
E. \textsc{Oset}$^1$, R.  \textsc{Molina}$^1$, D. \textsc{Nicmorus}$^2$, L.S.  \textsc{Geng}$^3$, P. \textsc{Gonzalez}$^1$, J. \textsc{Vijande}$^4$, S. \textsc{Sarkar}$^5$, 
Bao Xi \textsc{Sun}$^6$, M. J.  \textsc{Vicente Vacas}$^1$, A. \textsc{Ramos}$^7$,
E. J. \textsc{Garzon}$^1$, A. \textsc{Martinez Torres}$^8$, K. \textsc{Khemchandani}$^9$
}

\inst{$^1$Departamento de F\'{\i}sica Te\'orica and IFIC, Universidad de
Valencia-CSIC, E-46071 Valencia, Spain\\
$^2$Fachbereich Theoretische Physik, Institut f\"ur Physik, Karl-Franzens-Universitaet Graz, Universitaetsplatz 5, A-8010 Graz, Austria\\
$^3$Physik-Department, Technische Universitaet Muenchen,
D-85747 Garching, Germany\\
$^4$Departamento de F\'{\i}sica Atomica Molecular y Nuclear and IFIC,
Centro Mixto Universidad de Valencia-CSIC\\
$^5$Variable Energy Cyclotron Centre, 1/AF, Bidhannagar, Kolkata 700064, India\\
$^6$Institute of Theoretical Physics, College of Applied Sciences,
Beijing University of Technology, Beijing 100124, China\\
$^7$Departament d'Estructura i
  Constituents de la Mat\`eria and Institut de Ci\`encies del Cosmos,
  Universitat de Barcelona, 08028 Barcelona, Spain\\
$^8$Yukawa Institute for Theoretical Physics, Kyoto University, Kyoto 606-8502, Japan\\
$^9$Research Center for Nuclear Physics (RCNP), Osaka University, Ibaraki, Osaka 567-0047, Japan
}

\abst{   We present recent results on the vector meson-vector meson and vector meson-baryon interaction using 
a unitary approach based on the hidden-gauge Lagrangians. For the vector-vector case we find that 11 states get dynamically generated, corresponding to poles of the scattering matrices on
the second Riemann sheet. In the vector-baryon sector we also find 9 states dynamically generated from the vector-baryon octet interaction and 10 from the vector-baryon decuplet interaction. We also report on baryon states found from the interaction of two mesons and a baryon. 
}

\begin{document}

\maketitle

\section{Introduction}
 Early in 1978, S. Weinberg made a great gift to hadron physics by showing that the dynamics of QCD in the hadron world can be addressed
at low energies by means of effective theories in which the building blocks 
are the ground state mesons and baryons  \cite{Weinberg:1978kz}. This idea has produced
tools to address the interaction of mesons or mesons and baryons, 
mainly through chiral Lagrangians, which have had a tremendous impact on our
understanding of the spectrum of meson and baryon resonances. We all accept that
ground states of mesons and baryons are made of $q \bar{q}$ or three $q$
respectively. Yet, the spectrum of excited hadronic states can be much richer. 

 The building blocks in these chiral theories are the low energy hadrons, 
  such as the proton and baryons of its 
 SU(3) octet. To these one adds also the decuplet of the $\Delta$, considered as
 spin realignments of the three quark ground state. 
The basic mesons are the pion and mesons of its octet, to which one also adds
the nonet of the $\rho$, which also corresponds to spin realignments of the
$q \bar{q}$ ground state .

What about meson and baryon resonances?
The logical answer is that they are excitations of the quarks, which is the
essence of quark models. Yet, although extensive theoretical efforts are spent by QCD motivated approaches which describe hadron resonances as excitations of quarks \cite{Holl:2005vu,Krassnigg:2009zh}, 
 a complete and reliable calculation of the entire excited hadron spectrum is still missing. 

It is interesting to recall basic facts from the baryon spectrum.
The first excited $N^*$ states are the  
$N^*(1440)$ $(1/2^+)$ and the  $N^*(1535)$ $(1/2^-)$.
In quark models these states require a quark excitation of about
500-600 MeV. If this is the case, one may think that it takes less energy to 
create one pion, or two (140-280 MeV). The question is whether they can be
bound or get trapped in a resonant state.
How do we know if this can occur?
We need dynamics, a potential for the interaction of mesons with
ground state baryons and then solve the Schroedinger equation in coupled channels
(or the Bethe Salpeter equation where the mesons are treated relativistically). 
 This information can be 
extracted from chiral Lagrangians: the effective
theory of QCD at low energies. This is the philosophy behind the idea of 
dynamically generated  baryons:
Many resonances are generated in this way, like the $1/2^-$ states
from meson baryon: $N^* (1535)$ \cite{Kaiser:1995eg}, two $\Lambda(1405)$ 
\cite{cola} or the $1/2^+$ states
from two mesons and a baryon, like the $N^*(1710)$ and others 
\cite{alberto1,alberto2}.

Similarly, the interaction of pseudoscalar mesons leads to a good description of the low lying scalar mesons $f_0(600)$, $f_0(980)$, and $a_0(980)$ \cite{ramonet,Kaiser:1998fi,rios,Nieves:1999bx}

Inspired by the success of the unitary chiral approach, a further extension has recently been taken to
study  the interaction between two vector mesons and between one vector meson and one baryon \cite{Molina:2008jw,Geng:2008gx,nagahiro,Gonzalez:2008pv,Sarkar:2009kx,
angelsvec}. The novelty is that instead of using interaction kernels provided by ChPT, one uses transition amplitudes provided by the hidden-gauge Lagrangians, which lead to
a suitable description of the interaction of vector mesons among themselves and of vector mesons with other mesons or baryons. Coupled-channel unitarity works in the same way as in the unitary chiral approach, but now the dynamics is provided by
the hidden-gauge Lagrangians~\cite{hidden1,hidden2,hidden3,hidden4}.  As shown by several recent works \cite{Branz:2009cv,MartinezTorres:2009uk,Geng:2009iw}, this combination seems to work very well.

In this talk, we give an overview of recent developments concerning the interaction of vector mesons among themselves and vector mesons with baryons using the unitary approach, where several resonances appear as dynamically generated states.

\section{Theoretical framework}
We follow the formalism of the hidden gauge interaction for vector mesons of
\cite{hidden1,hidden2,hidden3,hidden4} (see also \cite{hidekoroca} for a practical set of Feynman rules).
The  Lagrangian involving the interaction of
vector mesons amongst themselves is given by
\begin{equation}
{\cal L}_{III}=-\frac{1}{4}\langle V_{\mu \nu}V^{\mu\nu}\rangle \ ,
\label{lVV}
\end{equation}
where the symbol $\langle \rangle$ stands for the trace in the $SU(3)$ space
and $V_{\mu\nu}$ is given by
\begin{equation}
V_{\mu\nu}=\partial_{\mu} V_\nu -\partial_\nu V_\mu -ig[V_\mu,V_\nu]\ ,
\label{Vmunu}
\end{equation}
where  $g$ is  $g=\frac{M_V}{2f}$,
with $f=93$ MeV the pion decay constant.  The magnitude $V_\mu$ is the ordinary 
$SU(3)$ matrix of the vectors of the octet of the $\rho$

The lagrangian ${\cal L}_{III}$ gives rise to a contact term coming from
$[V_\mu,V_\nu][V_\mu,V_\nu]$, as well as to a three
vector vertex 
\begin{equation}
{\cal L}^{(c)}_{III}=\frac{g^2}{2}\langle V_\mu V_\nu V^\mu V^\nu-V_\nu V_\mu
V^\mu V^\nu\rangle\ ;~~{\cal L}^{(3V)}_{III}
=ig\langle (V^\mu\partial_\nu V_\mu -\partial_\nu V_\mu
V^\mu) V^\nu\rangle,
\label{lcont}
\end{equation}

In this case one finds an analogy to the coupling of vectors to
 pseudoscalars given in the same theory by
\begin{equation}
{\cal L}_{VPP}= -ig \langle [
P,\partial_{\nu}P]V^{\nu}\rangle \ ,
\label{lagrVpp}
\end{equation}
where $P$ is the SU(3) matrix of the pseudoscalar fields.

In a similar way, one obtains the Lagrangian for the coupling of vector mesons to
the baryon octet given by
\cite{Klingl:1997kf,Palomar:2002hk} \footnote{Correcting a misprint in
\cite{Klingl:1997kf}}
\begin{equation}
{\cal L}_{BBV} =
g\left( \langle \bar{B}\gamma_{\mu}[V^{\mu},B]\rangle +
\langle \bar{B}\gamma_{\mu}B \rangle \langle V^{\mu}\rangle \right)
\label{lagr82}
\end{equation}
where $B$ is now the ordinary SU(3) matrix of the baryon octet

\begin{figure}[ht!]
\begin{center}
\centerline{\includegraphics[width=0.4\textwidth]{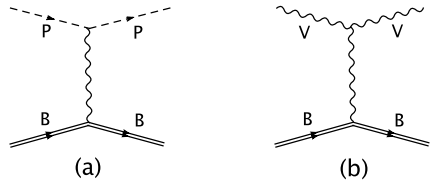}}
\caption{Diagrams contributing to the pseudoscalar-baryon (a) or vector-
baryon (b) interaction via the exchange of a vector meson.}
\label{fig:feyn}
\end{center}
\end{figure}

With these ingredients we can construct the Feynman diagrams that lead to the $PB
\to PB$ and $VB \to VB$ interaction, by exchanging a vector meson between the
pseudoscalar or the vector meson and the baryon, as depicted in Fig.
\ref{fig:feyn}.

From the diagram of Fig. \ref{fig:feyn}(a), and under the low energy approximation of
neglecting $q^2/M_V^2$ in the propagator of the exchanged vector, where $q$ is the
momentum transfer, one obtains the
same amplitudes as obtained from the ordinary chiral Lagrangian for
pseudoscalar-baryon octet interaction, namely the Weinberg-Tomozawa
terms.  The approximation of neglecting the three momenta of the vectors implies
that $V^{\nu}$ in eq. (\ref{lcont}) corresponds to the exchanged vector and the analogy
with eq. (\ref{lagrVpp}) is more apparent. Note that $\epsilon_\mu \epsilon^\mu$ becomes
$-\vec{\epsilon}\,\vec{\epsilon }\,^\prime$ and the signs of the Lagrangians also agree.

     One can see that the
     cases with $(I,S)=(3/2,0)$, $(2,-1)$ and $(3/2,-2)$, the last two
     corresponding to exotic channels, have a repulsive interaction
and do not produce poles in the scattering matrices.  However, the sectors
$(I,S)=(1/2,0)$, $(0,-1)$, $(1,-1)$ and $(1/2,-2)$ are attractive and one finds
 bound states and resonances in these cases.

    The scattering matrix is obtained  solving the
    coupled channels Bethe Salpeter equation in the on shell factorization approach of
    \cite{angels,ollerulf}
   \begin{equation}
T = [1 - V \, G]^{-1}\, V
\label{eq:Bethe}
\end{equation}
with $G$ being the loop function of a vector meson and a baryon. This function
is convoluted with the spectral function of the vector mesons to take into
account their width as done in \cite{nagahiro}.

 In this
case the factor $\vec{\epsilon}\,\vec{\epsilon }\,^\prime$, appearing in the potential $V$,
factorizes also in the $T$ matrix for the external vector mesons. This trivial
spin structure is responsible for having degenerate states with spin-parity
$1/2^-, 3/2^-$ for the interaction of vectors with the octet of baryons and
$1/2^-, 3/2^-, 5/2^-$ for the interaction of vectors with the decuplet
 of baryons.

  What we have done here for the interaction of vectors with the octet of
  baryons can be done for the interaction of vectors with the decuplet of
  baryons, and the interaction is obtained directly from that of the
  pseudoscalar-decuplet of baryons studied in
  \cite{kolodecu,Sarkar:2004jh}. The study of this interaction
  in \cite{Gonzalez:2008pv,Sarkar:2009kx,angelsvec} leads
  also to the generation of many resonances which are described below.

We search for poles in the scattering matrices in the second Riemann sheet, as
defined in previous works \cite{luisaxial}.

\begin{figure}[t]
\centerline{\includegraphics[scale=0.35]{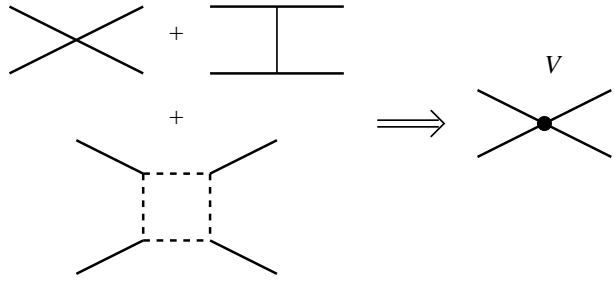}}
\caption{Transition amplitudes $V$ appearing in the coupled-channel Bethe-Salpeter equation.\label{fig:dia1}}
\end{figure}

For the case of vector-vector interaction the procedure followed is similar.
We outline the main ingredients of the unitary approach (details can be found in Refs.~\cite{Molina:2008jw,Geng:2008gx}). 
There are two basic building-blocks in this approach: transition amplitudes
provided by the hidden-gauge Lagrangians and a unitarization procedure. 
In Refs.~\cite{Molina:2008jw,Geng:2008gx}
three mechanisms,  as shown in Fig.~\ref{fig:dia1}, have been taken into account for the calculation of the transition amplitudes $V$:
the four-vector-contact term, the t(u)-channel vector-exchange amplitude, and the direct box amplitude with two intermediate pseudoscalar mesons. 

Among the three mechanisms considered for $V$, the
four-vector-contact term and $t(u)$-channel vector-exchange one are responsible for
the formation of resonances or bound states if the interaction generated by them
is strong enough. In this sense,
the dynamically generated states can be thought of as ``vector meson-vector meson molecules.'' On the
other hand, the consideration of the imaginary part of the direct box amplitude allows the generated states to decay into two pseudoscalars. It should
be stressed that in the present approach these two mechanisms play quite different roles:
the four-vector-contact term and the $t(u)$-channel vector-exchange one are responsible
for generating the resonances whereas the direct box amplitude mainly contributes to their decays.

 To calculate the box diagram, one has
to further introduce two parameters, $\Lambda$ and $\Lambda_b$. The parameter $\Lambda$ regulates the four-point loop function, and $\Lambda_b$ is related to the form factors used for the vector-pseudoscalar-pseudoscalar vertex, which
is inspired by the empirical form factors used in the study of vector-meson decays~\cite{Titov:2000bn}.
The values of $\Lambda$ and
$\Lambda_b$ have been fixed in Ref.~\cite{Molina:2008jw} to obtain the widths of the $f_0(1370)$ and $f_2(1270)$. They are found to provide also a good description of the widths of the $f'_2(1525)$, $K_2^*(1430)$, and $f_0(1710)$.

\section{Resonances from the interaction of two vector mesons}
\begin{table}[ht!]
      \renewcommand{\arraystretch}{1.4}
     \setlength{\tabcolsep}{0.1cm}
\caption{The properties, (mass, width) [in units of MeV], of the 11  dynamically
generated states and, if existing, of those of their PDG
counterparts~\cite{Amsler:2008zz}. The association of the dynamically generated states with their experimental counterparts is determined by matching their mass, width, and decay pattern.
\label{table:sum}}
\begin{center}
\begin{tabular}{c|c|cc|ccc}\hline\hline
$I^{G}(J^{PC})$&\multicolumn{3}{c|}{Theory} & \multicolumn{3}{c}{PDG data}\\\hline
              & Pole position &\multicolumn{2}{c|}{Real axis} & Name & Mass & Width  \\
              &               & $\Lambda_b=1.4$ GeV & $\Lambda_b=1.5$ GeV &           \\\hline
$0^+(0^{++})$ & (1512,51) & (1523,257) & (1517,396)& $f_0(1370)$ & 1200$\sim$1500 & 200$\sim$500\\
$0^+(0^{++})$ & (1726,28) & (1721,133) & (1717,151)& $f_0(1710)$ & $1724\pm7$ & $137\pm 8$\\
$0^-(1^{+-})$ & (1802,78) & \multicolumn{2}{c|} {(1802,49)}   & $h_1$\\
$0^+(2^{++})$ & (1275,2) & (1276,97) & (1275,111) & $f_2(1270)$ & $1275.1\pm1.2$ & $185.0^{+2.9}_{-2.4}$\\
$0^+(2^{++})$ & (1525,6) & (1525,45) &(1525,51) &$f_2'(1525)$ & $1525\pm5$ & $73^{+6}_{-5}$\\\hline
$1^-(0^{++})$    & (1780,133) & (1777,148) &(1777,172) & $a_0$\\
$1^+(1^{+-})$    & (1679,235) & \multicolumn{2}{c|}{(1703,188)} & $b_1$ \\
$1^-(2^{++})$    &  (1569,32) & (1567,47) & (1566,51)& $a_2(1700)??$
\\\hline
$1/2(0^+)$       &  (1643,47) & (1639,139) &(1637,162)&  $K_0^*$ \\
$1/2(1^+)$       & (1737,165) &  \multicolumn{2}{c|}{(1743,126)} & $K_1(1650)?$\\
$1/2(2^+)$       &  (1431,1) &(1431,56) & (1431,63) &$K_2^*(1430)$ & $1429\pm 1.4$ & $104\pm4$\\
 \hline\hline
    \end{tabular}
\end{center}
\end{table}

Searching for poles of the scattering matrix $T$ on the second Riemann sheet, we find
11 states in nine strangeness-isospin-spin channels as shown in Table I. Theoretical masses and widths are obtained with two
different methods: In the ``pole position'' method, 
the mass corresponds to the
real part of the pole position on the complex plane and the width corresponds to twice
its imaginary part. In this case, the box diagrams
corresponding to decays into two pseudoscalars are not included.
In the "real axis" method, the resonance parameters are
obtained from the modulus squared of the amplitudes of the dominant channel of each state
 on the real axis\footnote{See Tables I, II, and III of Ref.~\cite{Geng:2008gx}.}, where the mass corresponds to the energy at which the modulus squared has a maximum and the width corresponds to the difference
between the two energies where the modulus squared is half of the
maximum value. In this latter case, the box amplitudes are included.  The results
shown in Table I have been obtained using two different values of $\Lambda_b$,
which serve to quantify the uncertainties related to this parameter.

Our treatment of the box amplitudes enables us to obtain
the decay branching ratios of the generated states into two pseudoscalar mesons using the real-axis method and this is explained in detail in Ref.~\cite{Geng:2008gx}.

It is interesting to note that out of the 21 combinations of
strangeness, isospin and spin, we have found resonances only in nine of
them. In all the ``exotic'' channels, from the point of view that they cannot be formed from $q\bar{q}$
combinations, we do not find dynamically
generated resonances.

  Applications of the results exposed above have been done to study different processes. As an example, the radiative decay into $\gamma \gamma$ of the $f_2(1270)$ and $f_0(1370)$ has been studied in \cite{junkogam} and good agreement with experiment is found. Similarly, the decay into $\gamma \gamma$ or a $\gamma$ and a vector meson for other resonances of \cite{Geng:2008gx} has also been evaluated in \cite{Branz:2009cv}, with also good agreement with experiment when available. Another test successfully passed by the former approach to these resonances is the decay of $J/\psi$ into $\phi (\omega)$ and one of the resonances generated in \cite{Geng:2008gx}, as explained in \cite{MartinezTorres:2009uk}. Another test is the $J/\psi$ decay into a $\gamma$ and one of the resonances of \cite{Geng:2008gx}, which has been studied in 
\cite{Geng:2009iw}.

\section{Resonances from the interaction of vector mesons with baryons}

In table \ref{tab:pdg} we show a summary of the results obtained from the
interaction of vectors with the octet of baryons \cite{angelsvec} and the tentative
association to known states \cite{Amsler:2008zz}.

\begin{table}[ht!]
\caption{The properties of the 9 dynamically generated resonances and their possible PDG
counterparts.}\begin{center}
      \renewcommand{\arraystretch}{1.5}
     \setlength{\tabcolsep}{0.2cm}
\begin{tabular}{c|c|cc|ccccc}\hline\hline
$I,\,S$&\multicolumn{3}{c|}{Theory} & \multicolumn{5}{c}{PDG data}\\
\hline
    \vspace*{-0.3cm}
    & pole position    & \multicolumn{2}{c|}{real axis} &  &  & &  &  \\
    &   & mass & width &name & $J^P$ & status & mass & width \\
    \hline
$1/2,0$ & --- & 1696  & 92  & $N(1650)$ & $1/2^-$ & $\star\star\star\star$ & 1645-1670
& 145-185\\
  &      &       &     & $N(1700)$ & $3/2^-$ & $\star\star\star$ &
	1650-1750 & 50-150\\
       & $1977 + {\rm i} 53$  & 1972  & 64  & $N(2080)$ & $3/2^-$ & $\star\star$ & $\approx 2080$
& 180-450 \\	
   &     &       &     & $N(2090)$ & $1/2^-$ & $\star$ &
 $\approx 2090$ & 100-400 \\
 \hline
$0,-1$ & $1784 + {\rm i} 4$ & 1783  & 9  & $\Lambda(1690)$ & $3/2^-$ & $\star\star\star\star$ &
1685-1695 & 50-70 \\
  &       &       &    & $\Lambda(1800)$ & $1/2^-$ & $\star\star\star$ &
1720-1850 & 200-400 \\
       & $1907 + {\rm i} 70$ & 1900  & 54  & $\Lambda(2000)$ & $?^?$ & $\star$ & $\approx 2000$
& 73-240\\
       & $2158 + {\rm i} 13$ & 2158  & 23  &  &  &  & & \\
       \hline
$1,-1$ &  ---  & 1830  & 42  & $\Sigma(1750)$ & $1/2^-$ & $\star\star\star$ &
1730-1800 & 60-160 \\
  & ---    & 1987  & 240  & $\Sigma(1940)$ & $3/2^-$ & $\star\star\star$ & 1900-1950
& 150-300\\
   &     &       &   & $\Sigma(2000)$ & $1/2^-$ & $\star$ &
$\approx 2000$ & 100-450 \\\hline
$1/2,-2$ & $2039 + {\rm i} 67$ & 2039  & 64  & $\Xi(1950)$ & $?^?$ & $\star\star\star$ &
$1950\pm15$ & $60\pm 20$ \\
         & $2083 + {\rm i} 31 $ &  2077     & 29  &  $\Xi(2120)$ & $?^?$ & $\star$ &
$\approx 2120$ & 25  \\
 \hline\hline
    \end{tabular}

\label{tab:pdg}
\end{center}
\end{table}

  For the $(I,S)=(1/2,0)$ $N^*$  states there is the $N^*(1700)$ with
 $J^P=3/2^-$, which could correspond to the state we find with the same quantum
 numbers around the same energy. We also find in the PDG the  $N^*(1650)$, which
 could be the near degenerate spin parter of the $N^*(1700)$ that we predict in
 the theory. It is interesting to recall that in the study of
 Ref.~\cite{mishajuelich} a pole is found around 1700 MeV,
with the largest coupling to $\rho N$ states.
Around 2000 MeV, where we find another $N^*$ resonance,
there are the states $N^*(2080)$ and $N^*(2090)$, with $J^P=3/2^-$ and
$J^P=1/2^-$ respectively, showing a good approximate spin degeneracy.

For the case $(I,S)=(0,-1)$ there is in the PDG one state, the $\Lambda(1800)$
with $J^P=1/2^-$, remarkably close to the energy where we find a $\Lambda$
state.  The state obtained around 1900 MeV could
correspond to the $\Lambda(2000)$ cataloged in the PDG with unknown spin and parity.

 The case of the $\Sigma $ states having $(I,S)=(1,-1)$ is rather interesting.
 The state
that we find around 1830 MeV, could be associated to the  $\Sigma(1750)$
with $J^P=1/2^-$. More interesting seems to be the case of the state obtained around
1990 MeV that could be related to two PDG candidates, again
nearly degenerate, the $\Sigma(1940)$ and the $\Sigma(2000)$, with spin and
parity  $J^P=3/2^-$ and $J^P=1/2^-$ respectively.

  Finally, for the case of the cascade resonances, $(I,S)=(1/2,-2)$, we find
  two states, one  around 2040 MeV and the other one around 2080 MeV. There are two cascade states in
  the PDG around this energy region with spin parity unknown, the
  $\Xi(1950)$ and the $\Xi(2120)$.  Although the experimental
  knowledge of this sector is relatively poor, a program is presently running at
  Jefferson Lab to improve on this situation \cite{Nefkens:2006bc}.

    The case of the vector interaction with the decuplet is
similar \cite{Sarkar:2004jh} and we show the results in Table \ref{tab:pdg2}

\begin{table}[!ht]
      \renewcommand{\arraystretch}{1.5}
     \setlength{\tabcolsep}{0.2cm}
     \caption{The properties of the 10 dynamically generated resonances and their possible PDG
counterparts. We also include the $N^*$ bump around 2270 MeV and the $\Delta^*$ bump around 2200 MeV. }
\begin{center}
\begin{tabular}{c|l|cc|lcclc}\hline\hline
$S,\,I$&\multicolumn{3}{c|}{Theory} & \multicolumn{5}{c}{PDG data}\\\hline
        & pole position &\multicolumn{2}{c|}{real axis} & name & $J^P$ & status & mass & width \\
        &               & mass & width & \\\hline
$0,1/2$ & $1850+i5$   & 1850  & 11  & $N(2090)$ & $1/2^-$ & $\star$ & 1880-2180 & 95-414\\
        &             &       &     & $N(2080)$ & $3/2^-$ & $\star\star$ & 1804-2081 & 180-450\\
        &       &  $2270(bump)$ &  & $N(2200)$ & $5/2^-$ & $\star\star$ & 1900-2228 & 130-400\\
\hline
$0,3/2$ & $1972+i49$  & 1971  & 52  & $\De(1900)$ & $1/2^-$ & $\star\star$ & 1850-1950 & 140-240 \\
    &             &       &     & $\De(1940)$ & $3/2^-$ & $\star$ & 1940-2057 & 198-460   \\
        &             &       &     & $\De(1930)$ & $5/2^-$ & $\star\star\star$ & 1900-2020  & 220-500   \\
    &             & $2200 (bump)$  &     & $\De(2150)$ & $1/2^-$ & $\star$ & 2050-2200  & 120-200  \\
\hline
$-1,0$  & $2052+i10$  & 2050  & 19  & $\Lambda(2000)$ & $?^?$ & $\star$  & 1935-2030 & 73-180\\
\hline
$-1,1$  & $1987+i1$   & 1985  & 10   & $\Sigma(1940)$ & $3/2^-$  & $\star\star\star$ &
1900-1950 & 150-300 \\
        & $2145+i58$  & 2144  & 57  & $\Sigma(2000)$ & $1/2^-$  & $\star$ & 1944-2004 &
    116-413\\
    & $2383+i73$  & 2370  & 99 & $\Sigma(2250)$ & $?^?$ & $\star\star\star$ & 2210-2280 &
    60-150\\
    &   &   &  & $\Sigma(2455)$ & $?^?$ & $\star\star$ & 2455$\pm$10 &
    100-140\\
\hline
$-2,1/2$ & $2214+i4$  & 2215  & 9  & $\Xi(2250)$ & $?^?$ & $\star\star$ & 2189-2295 & 30-130\\
     & $2305+i66$ & 2308  & 66 & $\Xi(2370)$ & $?^?$ & $\star\star$ & 2356-2392 & 75-80 \\
         & $2522+i38$ & 2512  & 60  & $\Xi(2500)$ & $?^?$ & $\star$ & 2430-2505 & 59-150\\
\hline
$-3,0$   & $2449+i7$   & 2445 & 13  & $\Omega(2470)$   & $?^?$   & $\star\star$ & 2474$\pm$12 & 72$\pm$33\\
 \hline\hline
    \end{tabular}

\label{tab:pdg2}
\end{center}
\end{table}

 We also can see that in many cases the experiment shows the near degeneracy
 predicted by the theory. Particularly, the case of the three $\Delta$
 resonances around 1920 MeV is very interesting. One observes a near
 degeneracy in the three spins $1/2^-, 3/2^-, 5/2^-$, as the theory predicts. It
 is also very instructive to recall that the case of the  $\Delta(5/2^-)$ is
 highly problematic in quark models since it has a  $3~h\omega$ excitation
 and comes out always with a very high mass \cite{Gonzalez:2008pv,pedro}.

  The association of states found to some resonances reported in the PDG
  for the case of $\Lambda$, $\Sigma $ and $\Xi$ states looks also equally
  appealing as one can see from the table.

  The reasonable results reported here
   should give a stimulus to
  search experimentally for the missing spin partners of the already observed
  states, as well as possible new ones.

   One interesting application of these results is the evaluation of the radiative decay of these resonances into a $\gamma$ and a baryon of the octet or the decuplet, according to the case studied above. Predictions are made in 
\cite{Sun:2010bk} for radiative decay widths into $\gamma$ and a baryon, and, in particular, for the case of decay into $\gamma$ and a baryon of the octet, the helicity amplitudes are evaluated and compared to data when available.
 The agreement with experiment is fair, account taken of the experimental uncertainties. A comparison with predictions of quark models is also made in \cite{Sun:2010bk}.

\section{States of two mesons and a baryon}
 
In \cite{alberto1,alberto2} a formalism was developed to study Faddeev equations
of systems of two mesons and a stable baryon. The interaction of the pairs
was obtained from the chiral unitary approach, which proves quite successful to
give the scattering amplitudes of meson-meson and meson-baryon systems in the
region of energies of interest to us. The spectacular finding is that,
leaving apart the Roper resonance, whose structure is far more elaborate than
originally thought \cite{Krehl:1999km,Dillig:2004rh}, all the low lying
$J^P=1/2^+$ excited states are obtained as bound states or resonances of two
mesons and one baryon in coupled channels.

\begin{figure}[ht!]
\centerline{\includegraphics[scale=0.7]{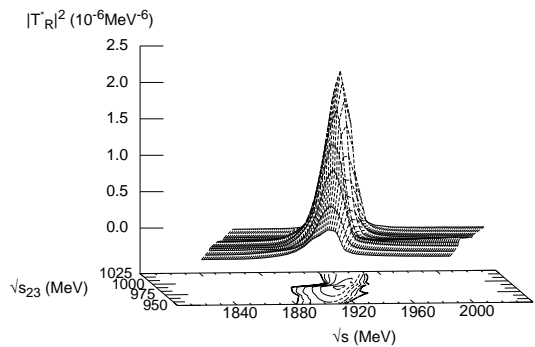}}
\caption{$|T|^2$ for the $N K\bar{K} \to N K\bar{K}$ transition, showing a peak that is associated to a tentative $N^*(1910)$ resonance.}
\label{threebody}
\end{figure}

  Particularly relevant is the issue of a possible bound 
state of $K \bar{K} N$. In \cite{jidothree},
using variational methods, the authors found a bound state of $K \bar{K} N$, with
the $K \bar{K}$ being in the $a_0(980)$ state \cite{jidothree}.  The system
was studied a posteriori in \cite{albertopheno} and it was found to
appear at the same energy and the same configuration, although with a mixture
of $f_0(980) N$, see fig. \ref{threebody}. This state appears around 1920 MeV with $J^P=1/2^+$. In a
recent paper \cite{albertoulf} some arguments were given to associate this state
with the bump that one sees in the $\gamma p \to K^+ \Lambda$ reaction around
this energy, which is clearly visible in recent accurate experiments
\cite{Bradford:2005pt,Sumihama:2005er}. If this association was correct there
would be other  experimental consequences, as an enhanced strength of the
$\gamma p \to K^+ K^- p$ cross section close to threshold, as well as a shift
of strength close to the $K \bar{K}$ threshold in the invariant mass
distribution of the kaon pair.  This reaction is right now under study 
\cite{nakanotalk}. Another
suggestion of \cite{albertoulf} is to measure the total $\gamma p$ spin 
$S_z=1/2$ and $S_z=3/2$
amplitudes, the $z$-direction along the photon momentum, since this 
would discriminate the cases where the peak around 
1920 MeV is due to a $1/2^+$ or a $3/2^+$ resonance.

%
\section{Acknowledgments}
L. S. Geng acknowledges support from the MICINN in the Program Juan de la Cierva. This work is partly
supported by DGICYT Contract No. FIS2006-03438, the Generalitat Valenciana in the program PROMETEO and the EU Integrated
Infrastructure Initiative Hadron Physics Project under contract
RII3-CT-2004-506078.


\begin{thebibliography}{99}
\bibitem{Weinberg:1978kz}
  S.~Weinberg,
  Physica A {\bf 96}, 327 (1979).
  
\bibitem{Holl:2005vu}
  A.~Holl, A.~Krassnigg, P.~Maris, C.~D.~Roberts and S.~V.~Wright,
  Phys.\ Rev.\  C {\bf 71}, 065204 (2005)
  [arXiv:nucl-th/0503043].

\bibitem{Krassnigg:2009zh}
  A.~Krassnigg,
  Phys.\ Rev.\  D {\bf 80}, 114010 (2009)
  [arXiv:0909.4016 [hep-ph]].


  
\bibitem{Kaiser:1995eg}
  N.~Kaiser, P.~B.~Siegel and W.~Weise,
  Nucl.\ Phys.\  A {\bf 594}, 325 (1995)  
  
 

  
\bibitem{cola}
  D.~Jido, J.~A.~Oller, E.~Oset, A.~Ramos and U.~G.~Meissner,
  Nucl.\ Phys.\  A {\bf 725}, 181 (2003)
  
\bibitem{alberto1}
  A.~Martinez Torres, K.~P.~Khemchandani and E.~Oset,
  Phys.\ Rev.\  C {\bf 77}, 042203 (2008)
  .


\bibitem{alberto2}
  K.~P.~Khemchandani, A.~Martinez Torres and E.~Oset,
  Eur.\ Phys.\ J.\  A {\bf 37}, 233 (2008)
 
 
  
\bibitem{ramonet}
  J.~A.~Oller, E.~Oset and J.~R.~Pelaez,
  Phys.\ Rev.\  D {\bf 59}, 074001 (1999)
  [Erratum-ibid.\  D {\bf 60}, 099906 (1999\ ERRAT,D75,099903.2007)]


\bibitem{Kaiser:1998fi}
  N.~Kaiser,
  Eur.\ Phys.\ J.\  A {\bf 3}, 307 (1998).
 



  
\bibitem{rios}
  J.~R.~Pelaez and G.~Rios,
  Phys.\ Rev.\ Lett.\  {\bf 97}, 242002 (2006)

\bibitem{Nieves:1999bx}
  J.~Nieves and E.~Ruiz Arriola,
  Nucl.\ Phys.\  A {\bf 679}, 57 (2000)
  
 
\bibitem{Molina:2008jw}
  R.~Molina, D.~Nicmorus and E.~Oset,
  Phys.\ Rev.\  D {\bf 78}, 114018 (2008)

  
\bibitem{Geng:2008gx}
  L.~S.~Geng and E.~Oset,
   Phys.\ Rev.\  D {\bf 79}, 074009 (2009).
  
  \bibitem{nagahiro}
  R.~Molina, H.~Nagahiro, A.~Hosaka and E.~Oset,
  Phys.\ Rev.\  D {\bf 80}, 014025 (2009)
 

\bibitem{Gonzalez:2008pv}
  P.~Gonzalez, E.~Oset and J.~Vijande,
  Phys.\ Rev.\  C {\bf 79}, 025209 (2009).


  
\bibitem{Sarkar:2009kx}
  S.~Sarkar, B.~X.~Sun, E.~Oset and M.~J.~V.~Vacas,
  Eur.\ Phys.\ J.\  A {\bf 44}, 431 (2010).

  
  


\bibitem{angelsvec}
  E.~Oset and A.~Ramos,
  Eur.\ Phys.\ J.\  A {\bf 44}, 445 (2010)
  [arXiv:0905.0973 [hep-ph]].

  

\bibitem{hidden1}
  M.~Bando et al. 
  Phys.\ Rev.\ Lett.\  {\bf 54}, 1215 (1985).
  
\bibitem{hidden2}
  M.~Bando, T.~Kugo and K.~Yamawaki,
  Phys.\ Rept.\  {\bf 164}, 217 (1988).
  
\bibitem{hidden3}
  M.~Harada and K.~Yamawaki,
  Phys.\ Rept.\  {\bf 381}, 1 (2003)
  
\bibitem{hidden4}
  U.~G.~Meissner,
  Phys.\ Rept.\  {\bf 161}, 213 (1988).
 
  


\bibitem{Branz:2009cv}
  T.~Branz, L.~S.~Geng and E.~Oset,
  Phys.\ Rev.\  D {\bf 81}, 054037 (2010)

\bibitem{MartinezTorres:2009uk}
  A.~Martinez Torres, L.~S.~Geng, L.~R.~Dai, B.~X.~Sun, E.~Oset and B.~S.~Zou,
  Phys.\ Lett.\  B {\bf 680}, 310 (2009)
  
\bibitem{Geng:2009iw}
  L.~S.~Geng, F.~K.~Guo, C.~Hanhart, R.~Molina, E.~Oset and B.~S.~Zou,
  Eur.\ Phys.\ J.\  A {\bf 44}, 305 (2010)


   

  

  
\bibitem{hidekoroca}
  H.~Nagahiro, L.~Roca, A.~Hosaka and E.~Oset,
  Phys.\ Rev.\  D {\bf 79}, 014015 (2009)
  .
  
  
\bibitem{Klingl:1997kf}
F.~Klingl, N.~Kaiser and W.~Weise,
Nucl.\ Phys.\ A {\bf 624} (1997) 527
.

\bibitem{Palomar:2002hk}
  J.~E.~Palomar and E.~Oset,
  Nucl.\ Phys.\  A {\bf 716}, 169 (2003)
  .
  
\bibitem{angels}
  E.~Oset and A.~Ramos,
  Nucl.\ Phys.\  A {\bf 635}, 99 (1998)
  
  
\bibitem{ollerulf}
  J.~A.~Oller and U.~G.~Meissner,
  Phys.\ Lett.\  B {\bf 500}, 263 (2001)

 
  
  
  
   

  
  
 


\bibitem{kolodecu}
  E.~E.~Kolomeitsev and M.~F.~M.~Lutz,
  Phys.\ Lett.\  B {\bf 585} (2004) 243
.




\bibitem{Sarkar:2004jh}
  S.~Sarkar, E.~Oset and M.~J.~Vicente Vacas,
  Nucl.\ Phys.\  A {\bf 750}, 294 (2005)
  [Erratum-ibid.\  A {\bf 780}, 78 (2006)]
  .
  

  
\bibitem{luisaxial}
  L.~Roca, E.~Oset and J.~Singh,
  Phys.\ Rev.\  D {\bf 72}, 014002 (2005)
  .

\bibitem{Titov:2000bn}
  A.~I.~Titov, B.~Kampfer and B.~L.~Reznik,
  Eur.\ Phys.\ J.\  A {\bf 7}, 543 (2000);
  A.~I.~Titov, B.~Kampfer and B.~L.~Reznik,
  Phys.\ Rev.\  C {\bf 65}, 065202 (2002).
  
\bibitem{Amsler:2008zz}
  C.~Amsler {\it et al.}  [Particle Data Group],
  Phys.\ Lett.\  B {\bf 667}, 1 (2008).
  
\bibitem{junkogam}
  H.~Nagahiro, J.~Yamagata-Sekihara, E.~Oset, S.~Hirenzaki and R.~Molina,
  Phys.\ Rev.\  D {\bf 79}, 114023 (2009)
 
\bibitem{mishajuelich}
  M.~Doring, C.~Hanhart, F.~Huang, S.~Krewald and U.~G.~Meissner,
  Nucl.\ Phys.\  A {\bf 829}, 170 (2009)




 
\bibitem{Nefkens:2006bc}
  B.~M.~K.~Nefkens,
  AIP Conf.\ Proc.\  {\bf 870}, 405 (2006).
  
\bibitem{pedro}
  P.~Gonzalez, J.~Vijande and A.~Valcarce,
  Phys.\ Rev.\  C {\bf 77}, 065213 (2008)
  .
  
\bibitem{Sun:2010bk}
  B.~X.~Sun, E.~J.~Garzon and E.~Oset, Phys. Rev. D, in print.
  arXiv:1003.4664 [hep-ph].
  
  
\bibitem{Krehl:1999km}
  O.~Krehl, C.~Hanhart, S.~Krewald and J.~Speth,
  Phys.\ Rev.\  C {\bf 62}, 025207 (2000)
  .

\bibitem{Dillig:2004rh}
  M.~Dillig and M.~Schott,
  Phys.\ Rev.\  C {\bf 75}, 067001 (2007)
  [Erratum-ibid.\  C {\bf 76}, 019903 (2007)]
  .
  

  
  

  
  
\bibitem{jidothree}
  Y.~Kanada-En'yo and D.~Jido,
  Phys.\ Rev.\  C {\bf 78}, 025212 (2008)
  .
  
  
\bibitem{albertopheno}
  A.~Martinez Torres, K.~P.~Khemchandani and E.~Oset,
  Phys.\ Rev.\  C {\bf 79}, 065207 (2009)

  
\bibitem{albertoulf}
  A.~Martinez Torres, K.~P.~Khemchandani, U.~G.~Meissner and E.~Oset,
  Eur.\ Phys.\ J.\  A {\bf 41}, 361 (2009)







\bibitem{Bradford:2005pt}
  R.~Bradford {\it et al.}  [CLAS Collaboration],
  Phys.\ Rev.\  C {\bf 73}, 035202 (2006)
  .

\bibitem{Sumihama:2005er}
  M.~Sumihama {\it et al.}  [LEPS Collaboration],
  Phys.\ Rev.\  C {\bf 73}, 035214 (2006)
  .
\bibitem{nakanotalk} T. Nakano, talk at the NSTAR2009 Workshop, Beijing, April
2009.
 




\end{thebibliography}
\end{document}